# Data Science for Social Good

**Ahmed Abbasi**
Human-centered Analytics Lab
Department of IT, Analytics, & Operations
University of Notre Dame
aabbasi@nd.edu

**Roger H. L. Chiang**
Department of Operations,
Business Analytics, & IS
University of Cincinnati
roger.chiang@uc.edu

**Jennifer J. Xu**
Department of Computer
Information Systems
Bentley University
jxu@bentley.edu

## Abstract

Data science has been described as the fourth paradigm for scientific discovery. The latest wave of data science research, pertaining to machine learning and artificial intelligence (AI), is growing exponentially and garnering millions of annual citations. However, this growth has been accompanied by a diminishing emphasis on social good challenges – our analysis reveals that the proportion of data science research focusing on social good is less than it has ever been. At the same time, the proliferation of machine learning and generative AI have sparked debates about the socio-technical prospects and challenges associated with data science; for human flourishing, organizations, and society. Against this backdrop, we present a framework for "data science for social good" (DSSG) research that considers the interplay between relevant data science research genres, social good challenges, and different levels of socio-technical abstraction. We perform an analysis of the literature to empirically demonstrate the paucity of work on DSSG in information systems (and other related disciplines) and highlight current impediments. We then use our proposed framework to introduce the articles appearing in the special issue. We hope that this article and the special issue will spur future DSSG research and help reverse the alarming trend across data science research over the past 30-plus years in which social good challenges are garnering proportionately less attention with each passing day.

## 1    Introduction

Data science is an interdisciplinary field that applies mathematics, statistics, machine learning, and data visualization techniques to extract insights and knowledge from data that are normally big and encompass both structured and unstructured formats. In March 2019, something extraordinary and unprecedented happened – an important milestone in the (relatively brief) history of data science. The three "godfathers" of deep learning – Geoff Hinton, Yoshua Bengio, and Yann LeCun – were awarded the prestigious 2018 Turing Award (Simonite, 2019). For those unfamiliar with the award, it is to computer science what the Nobel Prize is to disciplines such as economics and physics, or the Fields Medal to math. So why was it extraordinary? There are a couple of reasons. First, deep learning is essentially a class of machine learning methods (Samtani et al., 2023; Abbasi et al., 2016). If one were to look at prior seminal machine learning methods, none ever won the award[1]. Decision tree induction models (Quinlan, 1986) and their important extensions, such as random forests (Breiman, 2001), did not win despite being routinely ranked as the most used machine learning method for predictive analytics in research and practice over multiple decades (Abbasi et al., 2016). The same is true for support vector machines (SVM), which popularized the idea of learning problem/domain-specific representations and have been used extensively in prior information systems (IS) research (e.g., Abbasi et al., 2010; Chau et al., 2020). For both these methods, decision trees and SVMs, the seminal papers/authors have garnered over 200,000 citations on Google Scholar. On the other hand, as of the publication of this article, the deep learning Turing Award winners had amassed an astounding 1.5 million citations, underscoring the impact their work has had on research. Second, the depth of academic-industry engagement among the winners was unprecedented. At the time of the award, in addition to their academic positions at top universities in North America, all three had strong ties to major Silicon Valley tech companies[2].

---

[1] Judea Pearl won the Turing Award in 2011 for "contributions to artificial intelligence through the development of a calculus for probabilistic and causal reasoning" including Bayesian Networks, which have been used as a machine learning method for classification/prediction problems. However, his AI contributions are generally regarding as being broader than machine learning, whereas deep learning is widely regarded as a subset of machine learning.

The event signaled the culmination of a 30-year period in which the practice of data science, as well as academic research, has progressed as follows: data management, business intelligence (BI), statistical data mining and predictive analytics, and most recently, machine learning and artificial intelligence (AI) (Wixom & Watson, 2001; Chen et al., 2012; Agarwal & Dhar, 2014; Abbasi et al., 2016; Grover et al., 2018; Berente et al., 2021). Figure 1 illustrates this trend over the period 1990-2020 for three waves of data science research: data management & BI, data mining & analytics, and machine learning & AI. The y-axis depicts the number of new results (i.e., scholarly articles added) in the Google Scholar index for that given year across various data science topic keywords associated with the three waves. The figure shows the rise of data management and basic BI in the 1990s and the ascent of analytics in the 2000s. Being foundational to data science, both waves undoubtedly remain important and consequential today. Longitudinally, both follow the concave pattern commonly observed for many technology trends and research topics. In contrast, the figure also shows the exponential growth of machine learning and AI articles, generating an astounding 5 million results annually (fueling the aforementioned citation counts for foundational deep learning research).

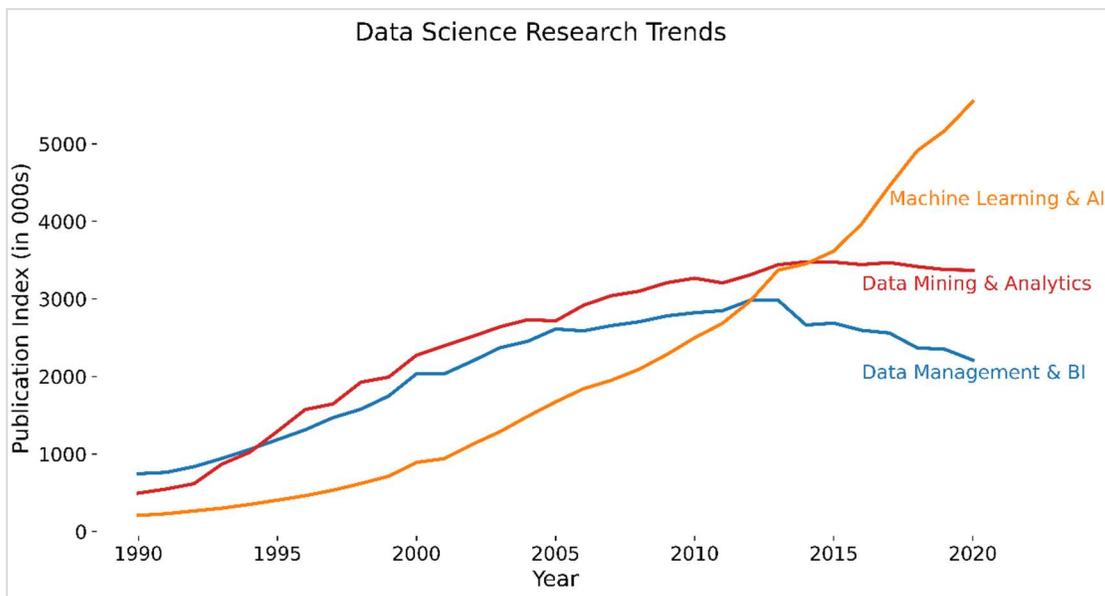

**Figure 1:** Google Scholar Index Trends for Three Waves of Data Science Research

In recent years, the proliferation of machine learning and generative AI have sparked debates about the prospects and challenges, most notably with the three deep learning Turing Award winners themselves split on the risks versus rewards[3]. On the one hand, data science in the age of advanced machine learning and deep learning presents tremendous opportunities for organizations and society. Historically, we have seen significant economic disruptions due to technology dating back to the agricultural and industrial revolutions. In the long run, these disruptions have improved the human condition. Opportunities abound for enhancing well-being and alleviating disparities. On the other hand, the risks and pitfalls are also very real. Given the socio-technical nature of modern data science research and applications, the central role of technology artifacts and technology firms in the data science revolution and evolution, and the broader need for thought leadership in an increasingly AI-enabled world, it is important to understand research opportunities for the field of information systems.

---

[2] At the time of the award, Hinton was a VP and Engineering Fellow at Google, LeCun was VP & Chief AI Scientist at Facebook, and Bengio was co-founder of Element AI, which was significantly backed by Microsoft's venture arm. Though not an award winner, at the time, Ian Goodfellow (Bengio's doctoral student) was Director of Machine Learning at Apple.
[3] In 2023, Geoff Hinton resigned from his position at Google to warn about the dangers of the technology. Bengio and LeCun were on opposing teams during a 2023 Munk Debate on Artificial Intelligence.

Jim Gray, a 1998 Turing Award winner, promoted data science as a new, fourth paradigm for scientific discovery in response to the large amounts of data generated by scientific experiments in many disciplines (Hey et al., 2009). In this vein, data science complements experimental, theoretical, and simulation science as an emerging research paradigm for understanding nature and society through "data-intensive computing" (Bell et al., 2009; Hey et al., 2009; p. xi). The inherently interdisciplinary nature of data science, and the fact that it is a catalyst for business transformation and technology disruption, presents many research opportunities for a diverse discipline such as IS. This has spurred a call for greater IS research on data science (Agarwal & Dhar, 2014; Saar-Tsechansky, 2015). Similarly, there is a need for research on the development and evaluation of data science artifacts (e.g., models, methods, and systems) that address broader societal challenges. A lingering question remains: what societal challenges can IS-oriented data science research contribute towards – and how can we conduct such research to maximize impact and relevance?

The purpose of the special issue was to explore the intersection of "Data Science for Social Good" (DSSG) from the perspective of IS. We limit our discussion to data science research as opposed to the practice of data science in industry, which has already received ample attention (e.g., Davenport & Patil, 2012). Accordingly, the goal of this article is three-fold. First, we present an IS-oriented framework for DSSG research that considers the interplay between relevant IS data science research genres, social good challenges (i.e., problems/outcomes), and different types of socio-technical abstractions. Second, we perform an analysis of the literature to empirically demonstrate the paucity of work on DSSG in IS, and other related disciplines, and highlight current impediments. Third, we use our proposed framework to introduce the articles appearing in the special issue.

## 2      Data Science and Social Good – A Motivating Example

Before delving into our proposed framework for DSSG, to further underscore the need for research at the intersection of data science and social good, it is only fitting that we use data science techniques to help understand the current state of data science research. Accordingly, we analyzed all research appearing in the Google Scholar index for the thirty-three-year period spanning 1990-2022. In our analysis, we used two sets of keywords: one related to data science topics (e.g., data mining, data visualization, data science, machine learning, deep learning) and another pertaining to social good challenges such as poverty, hunger, inequalities, clean water, sanitation, peace, justice, sustainable communities, and affordable clean energy (Cowls et al., 2021). For all combinations of keywords appearing within and across the two sets (e.g., "deep learning" and "poverty"), we gathered the number of articles appearing in Google Scholar annually for our analysis period. We then visualized the publication keyword co-occurrences using network analysis over three time periods: 1990-2000, 1990-2010, and 1990-2022.

Before conducting the analysis, we were certain that the longitudinal trend would be one where social good topics play a more prominent role in data science research over time. We thought that our key takeaways would relate to the slow rate of progress, that is, the trajectory. Or about the relatively nascent emerging role of IS in an increasingly vibrant DSSG landscape. What we observed – much to our chagrin – was a completely different panoramic picture. The results appear in Figure 2. In each time period panel, blue nodes signify data science-related keywords, and red nodes connote social good challenges. Consistent with co-occurrence network analysis, edge tie strengths between any two nodes were quantified as the percentage of articles containing *both* keywords, relative to the total number of articles appearing across the two keywords. A spring layout algorithm was used to arrange nodes based on their respective tie strengths such that keywords were arranged based on their percentage overlap with other keywords. Panel (a) shows that data science and social good topics are closely intertwined for the 1990-2000 period, as evidenced by the spatial mixture of blue and red nodes. For instance, *data mining* is closely connected to *health and wellness*, *clean water*, and *affordable energy*. *Data visualization* has strong ties with *sustainable communities*. *Analytics* is discussed in close relation with *well-being*. *Data science* lies near *poverty*. *Meta-learning* is closely connected with *hunger*. To those who have been in the data science space for over 20 years, this connection between data science and societally impactful application contexts is not overly surprising.

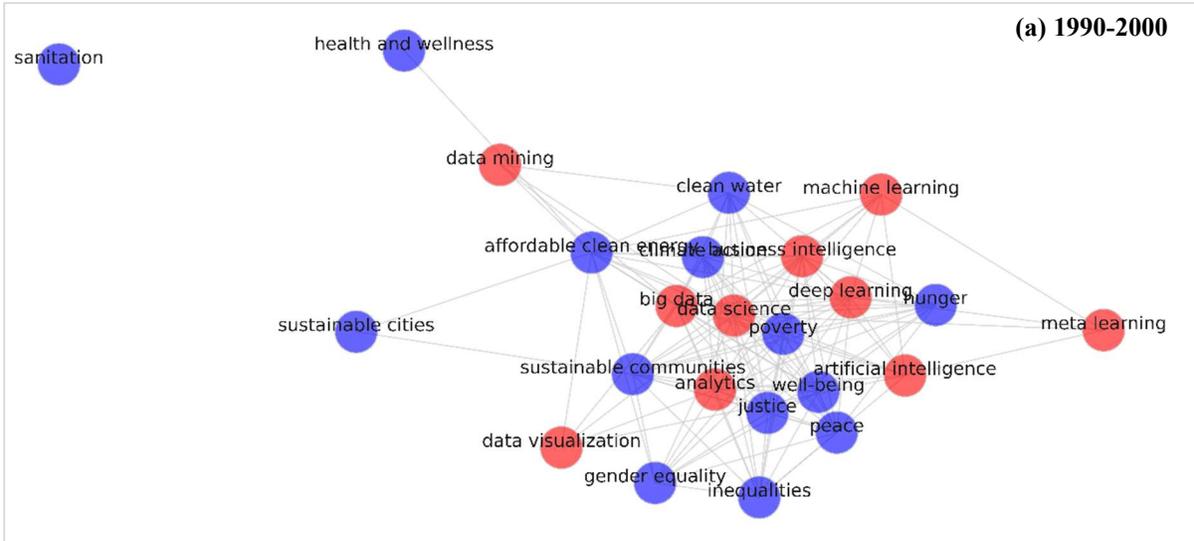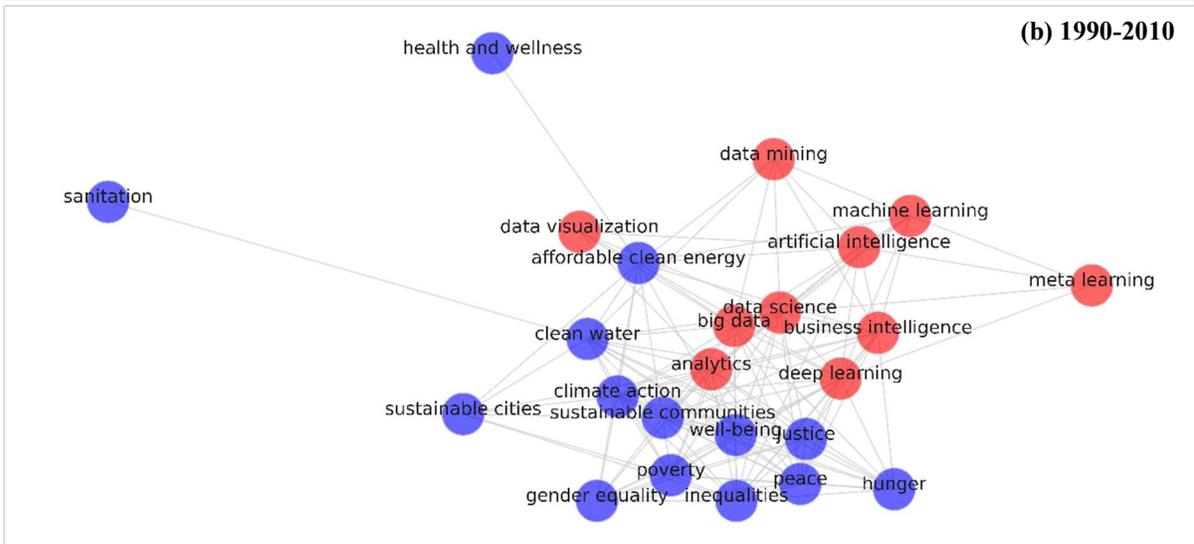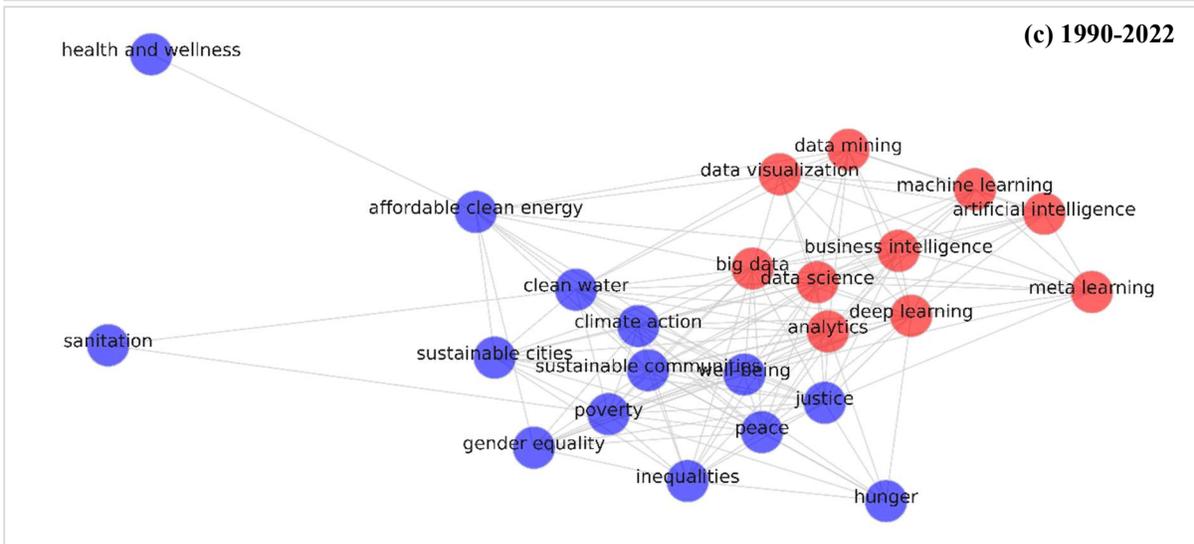

**Figure 2:** Co-occurrence Networks for Data Science Keywords (red nodes) and Social Good Topics (blue nodes) Across Three Time Periods Between 1990-2022

Looking at the co-occurrence network in the middle panel (b) in Figure 2, we see that the data science and social good topics are less intertwined for the 1990-2010 period. Although *affordable clean energy* maintains strong ties with keywords such as *data visualization* and *big data*, the two sets of keywords have drifted further apart. In the bottom panel (c) in Figure 2, we extend our analysis to include the entire body of research spanning 1990-2022. We observe that the two different node colors are further coalescing within themselves. Topics such as *sanitation* are now more closely discussed with *clean water* and *poverty*, suggesting that social good research is making connections amongst interrelated antecedent and/or consequent concepts. Regarding linkages across data science and social good keywords, *analytics* and *deep learning* maintain ties with *hunger*, *well-being*, and *justice*. However, the distances within keyword sets are reducing, whereas the distances across sets are increasing. This means the two bodies of research are clustering within, resulting in relatively less overlap. On the surface, this trend is somewhat surprising and counter-intuitive. Social good themes are undoubtedly garnering greater attention. However, our analysis does not focus on the amount of research in absolute terms – those numbers are increasing for data science *and* social good. Instead, we are focusing on the percentage overlap as a proportion of all research undertaken in the two areas. We can infer that the proportion of data science research geared towards social good themes has decreased over time[4].

One could argue that the absolute quantity of research is more important than proportionality. Furthermore, our analysis was intentionally simple in that we did not weight articles based on citation counts. Nonetheless, *proportion* or *share-of-volume does matter*. It is routinely used to examine dominant topics, diversity of research themes, and temporal dynamics and trends (Mustak et al., 2021). The current trend sends a strong signal regarding where the lion's share of time and effort pertaining to data science research is heading in relation to social good topics. Modern data science research is dominated by deep learning studies with goals such as the creation of "foundation models" that codify a vast array of knowledge, for instance, related to computer vision or language, with aspirations of artificial general intelligence (Bubeck et al. 2023). We cannot help but wonder whether data science has become so engrossed with generalizable, foundational methods research, guided by the common task framework (Liberman, 2010; Donoho 2017), that it has lost sight of the noble intents and purposes that made the promise of data science as a mechanism for good exciting in the first place.

## 3   A "Data Science for Social Good" Framework

### 3.1   Defining DSSG for IS

Data Science embodies a wide array of methods, including machine learning and statistics, applied to large quantities of structured and unstructured data. It has been defined in different ways, both in regards to its role in practice, and as a mechanism for conducting scientific research. For instance, Provost and Fawcett (2013, p.52) defined it as "…a set of fundamental principles that support and guide the principled extraction of information and knowledge from data" and "as the connective tissue between data-processing technologies…and data-driven decision making." Dhar (2013, p. 64) states, "Data science might therefore imply a focus involving data and, by extension, statistics, or the systematic study of the organization, properties, and analysis of data and its role in inference, including our confidence in the inference." Two things are apparent from these definitions: (1) data science research occurs within organizational, institutional, and/or societal environments; (2) the inferences and outcomes are closely tied to the underlying problem contexts. Notably, one challenge when talking about data science in general – similar to big data – is that the term can be very broad and widely applicable. Figure 3 presents a Venn diagram that captures these properties as they relate to DSSG in IS: the intersections of data science research genres, socio-technical environments, and social good challenges. In the remainder of the section, we discuss these in detail.

#### 3.1.1   Data Science Research Genres

*Computational social science* draws inferences about individuals or groups from large, longitudinal, digital data on human interactions (Lazer et al., 2009; Edelmann et al., 2020). Often-cited examples include deriving patterns and insights from large social networks (Aral & Walker, 2012; Agarwal & Dhar, 2014) or mining large language corpora (Kozlowski et al., 2019). Computational social science focuses on deriving important patterns and insights from structured and unstructured data, leading to empirical generalizations.

---

[4] We acknowledge that the uptick in social good themed conferences, special issues, and research on topics such as justice, poverty, clean water, and inequalities, as well as new schools/colleges focusing on urban analytics and climate and sustainability, will undoubtedly help reverse this trend. The rise of generative AI is ushering a new wave of ethics research on algorithmic bias, fairness, privacy, and related responsible AI themes.

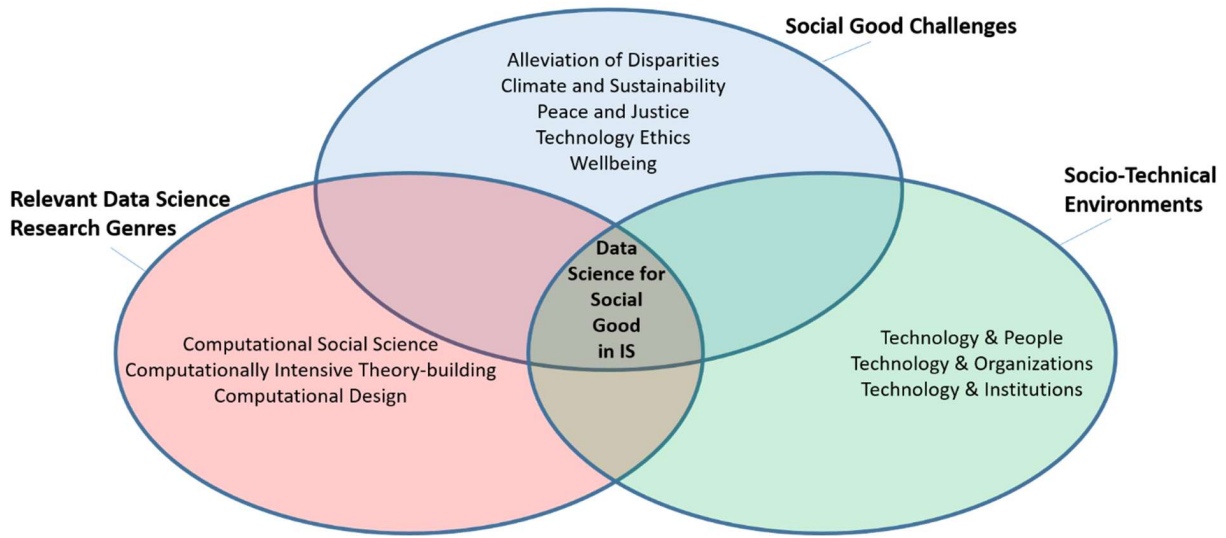

**Figure 3:** Data Science for Social Good (DSSG) in Information Systems (IS): The Intersection of Research Genres, Social Good Challenges, and Socio-Technical Environments

*Computationally intensive theory construction* builds on this idea, but specifically emphasizes theory construction and the necessary prerequisite guardrails and formalism (Berente et al., 2019; Miranda et al., 2022). The process of pattern surfacing involves consideration of the research communities' knowledge and cumulative tradition, manifesting in the form of lexical framing (Miranda et al., 2022). More specifically, lexical framing entails the use of practice lexicons (elements of the phenomena), method lexicons (computational techniques), and theoretical lexicons (concepts and associations).

*Computational design* is a genre of design science that emphasizes solving business and societal problems through the development of computational artifacts that are impactful, relevant, and often interdisciplinary (Rai, 2017). There has been a significant uptick in interest regarding machine learning-oriented computational design research (Padmanabhan et al., 2022). Computational design research relates to developing novel models or algorithms that offer design guidelines and/or best practices that contribute to a cumulative design tradition for a class of problems and/or artifacts.

It is important to state that the three research genres discussed do not necessarily have a one-to-one mapping with the data science paradigm. For instance, computational social science also belongs to the research genre taxonomy for social science. Computationally intensive theory construction relates to the grounded theory approach (Berente et al., 2019), and also encompasses the use of simulation methods (Miranda et al., 2022), which are often not associated with the data science paradigm (Hey et al., 2009). As alluded to, computational design is also a sub-genre of design science, and furthermore, relates to the technical branch of quantitative research alongside other quantitative branches such as analytical and empirical. It is also worth noting that the genres included are illustrative of data science, not exhaustive. For example, computational economics and machine learning-based econometrics could also be considered data science. Traditional econometrics and experimental methods for causal inference also undoubtedly have a role to play (and already are) regarding social good research. However, because we rely on Gray's definition of paradigms (Hey et al., 2009), we limit our discussion to the three aforementioned genres.

### 3.1.2  Social Good Challenges

Over the centuries, the term "social good" has been defined in many different ways. It has its roots in philosophy, where the idea of human flourishing through the "common good" traces back to Aristotle and refers to a good only attainable by the community, but shared by its members (MacIntyre, 1984; Smith, 1999). Notably, developing a list of what constitutes common good, as well as how to prioritize common good causes in resource-constrained environments, are both considered non-trivial due to a lack of consensus (Smith, 1999; Kraut, 2022). Some recent studies refer to it as "services or products that promote human well-being on a large scale" (Mor Barak, 2020, p. 139). Here, the products and services may refer to timely access to healthcare, clean water, quality education, equal rights, and so on (Mor Barak, 2020). Some social good studies have proposed the use of the United Nations Sustainable Development Goals (SDGs) – 17 goals related to alleviating poverty, hunger, and inequalities while promoting health and well-being, education, gender equality, clean water, affordable clean energy, climate action, sustainable cities, peace and justice, and life on land and water, etc. (Cowls et al., 2021). These can be distilled down to social good challenges related to alleviation of disparities, climate and sustainability, peace and justice, well-being, and, of course, technology ethics (including responsible AI).

### 3.1.3 Socio-technical Environments

IS research has long-standing traditions in exploring socio-technical phenomena across multiple levels of analysis. These include the individual, organizational, and institutional levels. The individual level of analysis focuses on how people adopt and use technologies in organizations and online environments (Davis, 1989; Goodhue & Thompson, 1995; Pavlou & Gefen, 2004). The organizational level emphasizes the successful design, implementation, and practices associated with technologies within traditional organizations, as well as in innovative forms such as virtual teams and online communities (Markus, 1983; DeLone & McClain, 1992; Orlikowski, 1992; Ren et al., 2012; Hirschheim et al., 1995). Finally, the institutional level of analysis refers to the societal environment of organizations and involves research into the diffusion and impact of technologies in industry, as well as in fields such as healthcare and governments, and in developing nations (Brynjolfsson, 1994; King et al., 1994; Swanson & Ramiller, 1997; Angst et al., 2010).

DSSG has a natural fit with IS for a couple of reasons. First, social good relates to humanistic outcomes – an important consideration for socio-technical IS work and complementary to research focusing on instrumental outcomes (Sarker et al., 2019). Second, some of the data science genres alluded to, namely computationally intensive theory construction and computational design, have strong roots in IS. For these reasons, we believe that future IS work on DSSG has an opportunity to align with the calls for differentiable research, "a unique perspective in order to identify, support, and/or legitimize research within that discipline" (Sarker et al., 2019; p. 699).

## 3.2 DSSG Framework and Examples

Figure 4 presents a framework based on the data science (DS) genre, social good (SG) challenges, and socio-technical environmental granularities (IS). What is notable is that data science can potentially create interesting intersections with social good (DSSG) and IS (DS in IS), as well as the intersection of all three, represented by the cube (DSSG in IS). What is intentionally missing in Figure 4 are guidelines for the *rigor* of the data science tools, techniques, and principles leveraged, *depth* of the social good contextualization in terms of immersion in field settings and downstream implications, and the *cohesion* of the socio-technical environments to the research motivation, undertaking, or outcomes. Readers interested in best practices on data science rigor can turn to various textbooks and editorials on the matter. Those interested in perspectives on the depth of social good contextualization can look at Cowls et al. (2021). We encourage readers interested in how to effectively engage with the socio-technical IS cumulative tradition to peruse Sarker et al. (2019). The framework emphasizes the DSSG and IS intersections that guide which cell a study or stream may fall under. In contrast, the rigor/depth/cohesion are collectively indicators of quality (cell agnostic) and centrality (cell-specific) within the DSSG for IS cube.

Table 1 presents examples of studies related to DSSG at different levels of socio-technical granularity (e.g., people, organizations, and institutions). The three rows relate to the DS genres, whereas the three columns depict socio-technical environments from an IS vantage point. Each cell features examples of studies that explore social good challenges using a particular DS genre and a certain socio-technical granularity level. We intentionally include studies appearing in top IS outlets, as well as those published by IS scholars in adjacent fields/journals, and also studies conducted in related fields by sociologists and computer scientists. In doing so, we believe the chosen studies are interesting examples of DSSG in terms of rigor and depth of engagement with social good challenges but are not necessarily exemplars of DSSG in IS in terms of cohesion with the IS socio-technical cumulative tradition. Because the DSSG in IS literature is emerging (we explicitly discuss this emerging space later in Section 4.1), our goal is to include a panorama of interesting DSSG work to inspire a new wave of DSSG research in IS.

In Table 1, the first row depicts studies from the computational social science genre. The mode for computational social science research examining social good challenges has been to analyze social networks with the objective of revealing node characteristics, structural linkage patterns, and/or information diffusion patterns over time across organizations or institutions (Lazer et al., 2009; Aral & Walker, 2012). For instance, Xu and Chen (2005) construct criminal activity networks for narcotics gangs in the southwestern United States responsible for methamphetamine trafficking. Others have examined the role of key actors in online social movement organizations that promote racial and religious prejudice and violence (Chau & Xu, 2007; Zimbra et al., 2010). At the institutional level, computational social science studies have explored susceptibility to influence and the propagation of fake news (Vosoughi et al., 2018), with dire implications for media institutions, digital platforms, and the role of technology in supporting the pursuit of evidence and truth (Lee & Ram, 2023).

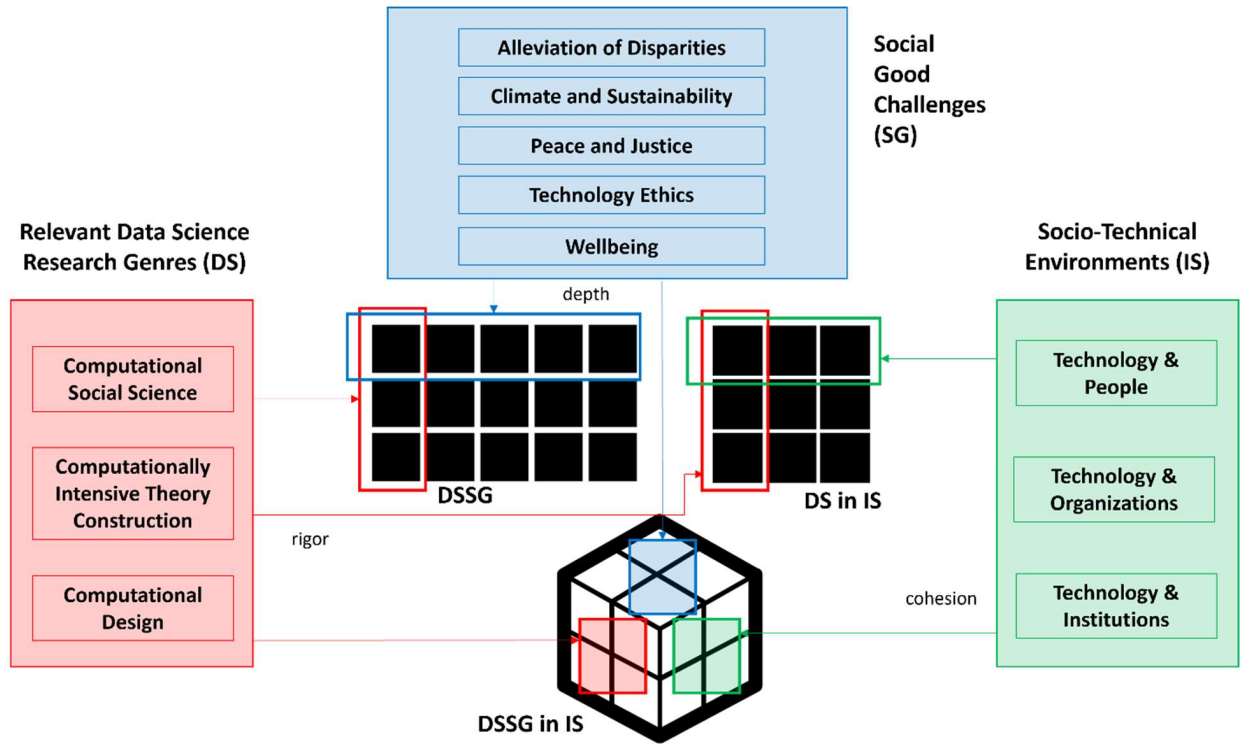

**Figure 4:** Data Science for Social Good (DSSG) in Information Systems (IS): A Framework that Considers Research Genres, Social Good Challenges and Contexts, and Socio-Technical Environments

Table 1: Data Science for Social Good (DSSG) in Information Systems (IS): Research Examples

|  | Technology & People | Technology & Organizations | Technology & Institutions |
|---|---|---|---|
| **Computational Social Science** | **Using word embeddings to understand gender/ethnic disparities** (Garg et al., 2018) | **Crime network analysis** (Xu and Chen 2005)<br><br>**Analyzing online hate communities** (Chau and Xu 2007)<br><br>**Analyzing social movement organizations** (Zimbra et al. 2010) | **Using LLMs to identify aspects of societal culture** (Kozlowski et al. 2019)<br><br>**Fairness in ratemaking** (Zhang and Xu 2024)<br><br>**Spread of true and false news** (Vosoughi et al. 2018) |
| **Computationally Intensive Theory Construction** | TBD | **Firm Communication During Disasters** (Yan et al. 2024) | TBD |
| **Computational Design** | **ML4H – Machine learning for health: fairness & disparities** (Sarkar et al. 2020)<br><br>**AI4SG – AI for social good** (Cowl et al. 2021)<br><br>**Proactively Detecting Emotional Distress** (Chau et al. 2020) | **Decision support for wildfire management** (Gomez et al. 2024) | **ML4D – Machine learning for the developing world** (De-Arteaga et al. 2018)<br><br>**Improving Drinking Water Access and Equity in Africa** (Zhai et al. 2023)<br><br>**Credibility by Design in Listening for Public Health 3.0** (Kitchens et al. 2024) |

An exciting development within the computational social science space is the use of text analysis to derive important insights into social good challenges. Garg et al. (2018) show that word embeddings, a natural language processing technique commonly used in modern data science research and practice, can learn and reflect gender and ethnicity-related stereotypes. Using data from 1910 to 1990, they find that gender biases in word embeddings trained on various available textual data sources from across the twentieth century are highly correlated with occupational differences between men and women from those time periods. At the institutional level, Kozlowski et al. (2019) use word embeddings to explore "the geometry of culture" as it relates to evolving markers of social class (e.g., wealth, education, etc.). The two studies illustrate how patterns codified in word embeddings and large language models (LLMs) can shed light on social constructs and cultural artifacts involving people, organizations, institutions, and society. They also highlight the dangers of machine learning models trained on historical data regarding the current and future role of AI-enabled systems in supporting social progress and a more just society (Kane et al., 2021).

The frequency of climate-related and man-made disasters has accelerated in recent years. Moreover, disasters often have a disproportionately greater impact on those most vulnerable. Within the computational social science genre applied to the institutional level, Zhang and Xu (2024) explore the fairness of ratemaking in catastrophe insurance. Using concepts from machine learning, they shed light on the disparate impact current ratemaking policies have on minorities. Given the nascent body of research on computationally intensive theory construction, we briefly discuss DSSG research related to this genre. Within the disaster response arena, Yan et al. (2024) propose a framework with a novel word embedding approach to explore the impact of message orientations of firm communication during disasters on public engagement. They use the emerging nature of disasters, and lack of well-established theories capable of offering strong support for formal deductive hypotheses, to motivate their use of a computationally intensive theory construction framework. Based on the dynamics of social good challenges, we envision ample opportunities for computationally intensive theory building within the DSSG in IS space.

Similar to computationally intensive theory construction, the computational design genre presents unique opportunities for IS to offer differentiable thought leadership in the DSSG in IS space. At the intersection of technology and people are efforts related to ML4H (machine learning for health) focused on "accessible diagnostic and prognostic systems, health equity, fairness and bias, generalization across populations or systems, improving patient participation in health…" (Sarkar et al., 2020; p. 2). Other examples include AI4SG (AI for social good), the use of AI to improve the human condition as it relates to the United Nation's sustainable development goals (Cowls et al., 2021), and the use of machine learning to identify online users battling emotional distress (Chau et al., 2020). At the organizational level, Gomez et al. (2024) tackle the problem of how wildfire management agencies can better budget for upcoming wildfire seasons through design artifacts that couple predictive and prescriptive models that consider climate uncertainty and opportunity costs of over/under-budgeting.

Computational design research considering institution-level characteristics and application scale represents an important and exciting opportunity for DSSG in IS. How can we identify the optimal arrangement of wells for drinking water in Ethiopia with limited digitized knowledge of resources, constraints, and village/community dynamics? This is the challenge Zhai et al. (2023) tackle, using field interviews and site visits to identify data inputs for prescriptive models, including the dynamics of adjacent communities (that often coordinate on water usage and transportation across neighboring villages), the impact of war, as well as efficiency and equity considerations. In the same vein, De-Arteaga et al. (2018, p. 3) discuss ML4D (machine learning for the developing world) scenarios where "existing or plausible solutions in developed regions are not viable" for developing countries. At the institutional level, examples include modeling patterns of violence and human rights violations and detecting corruption in international development contracts. Kitchens et al. (2024) explore the design of social listening platforms for problems such as opioid/drug epidemics in online settings rife with low-credibility content.

From our discussion of the DSSG framework, we are optimistic that opportunities abound for societally impactful IS research using data science. In the next section, we discuss the current business/management/IS research landscape related to DSSG and then offer high-level guidelines of what we believe is needed to foster greater DSSG research.

# 4  The Current State of DSSG Research

## 4.1  Meta-analysis

We performed a literature analysis of DSSG articles published over the 11-year time-period spanning 2012–2022 to provide an overview of how IS and other related disciplines have used data science-related computational genres to address societal challenges, and where these existing studies could be positioned in the DSSG framework. A summary of our analysis appears in Figure 5, including the genres, social good challenges, and socio-technical environments. We discuss the process undertaken before delving into the figure results and takeaways. Because the primary goal of this analysis was to be illustrative as opposed to exhaustive, we focused on nine IS journals, including the *Senior Scholars' Basket of Eight* and *Decision Support Systems*. To have a broader view of the DSSG research in other related disciplines, we expanded our literature survey to include Accounting, Finance, Management, and Economics journals in the *Financial Times Top 50* list and five Sociology journals identified by the Scimago Journal & Country Rank: *American Sociological Review*, *Annual Review of Sociology*, *Journal of Information Communication & Ethics in Society*, *Sociological Methods & Research*, and the *American Journal of Sociology*. We also included two interdisciplinary journals: *INFORMS Journal of Data Science* and *Management Science*.

To ensure the broadest coverage possible, we started the literature search using a single keyword: *data*. An article was included in the sample if the word "data" appeared in the article's title, abstract, subjects, or keyword list. The resulting set consisted of 6,920 data-related articles published in 42 journals. Two graduate research assistants read the titles and abstracts of these articles to filter out articles unrelated to data science or that involved non-computational research genres. Examples of excluded articles are editorials, research commentaries and opinions regarding big data and technologies, empirical articles not using big data or computational methods, etc. The filtered dataset comprised 659 data science articles, the majority of which (540 articles) addressed various business, economic, and/or instrumental objectives (e.g., revenue, productivity, efficiency). Figure 5 refers to these as the data science (DS) set of articles. Only 119 articles tackled social issues aligned with our broad list of social good challenges. Figure 5 refers to these as the DSSG set.

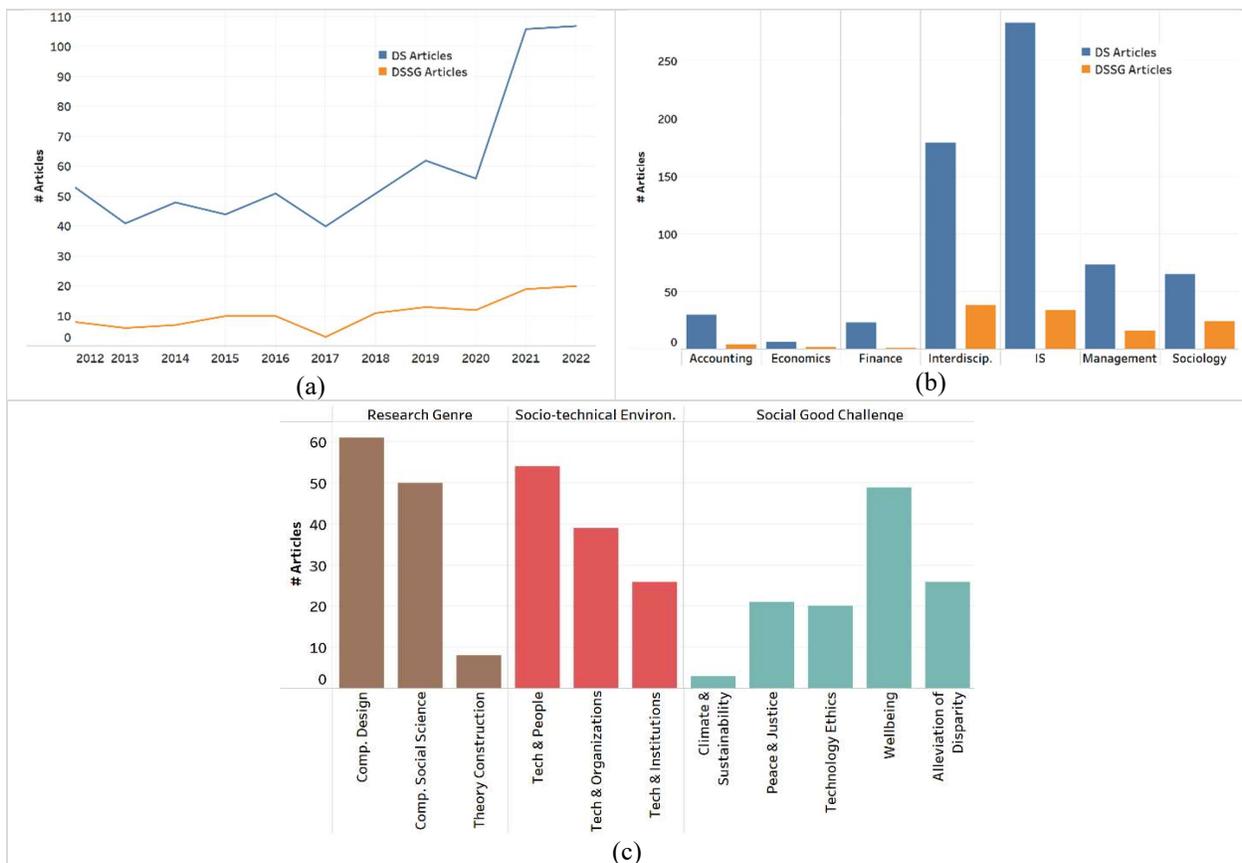

**Figure 5**: Number of DS articles and DSSG articles published during 2012-2022, depicted by year (a), by discipline (b), and by genre, social good challenge, and socio-technical environment (c).

Figure 5a presents changes in the counts of DS and DSSG articles by year. This includes a sizable jump in the number of published DS articles in 2021 and onwards. However, although the number of DSSG articles per year has doubled, the proportion focusing on social good has not increased. Looking at the breakdown of DS and DSSG articles by disciplines (Figure 5b), we find that the nine IS journals have the highest number of data science articles published, followed by the two interdisciplinary journals, Management, and Sociology journals. Journals in other disciplines, such as Finance, Accounting, and Economics, have markedly fewer DS articles. As mentioned above, because the majority of DS papers in these outlets address instrumental business problems, the number of published DSSG articles is significantly lower (under 20% of all DS research).

In addition, based on our proposed DSSG research framework, we hand-coded the data science research genres, social good challenges, and socio-technical environments for the 119 DSSG articles identified through our systematic review (these appear in Figure 5c). In terms of the research genres, the vast majority of DSSG articles are devoted to the design of computational artifacts or computational approaches to uncover social science insights. Conversely, only a few articles seek to surface computationally-derived patterns with theory-building implications related to social good. This is intuitive given the relatively longer cumulative traditions associated with computational design (e.g., Arazy & Woo, 2007; Abbasi & Chen, 2008) and computational social science (Lazer et al., 2009). Regarding social good challenges, many articles investigate challenges related to well-being (e.g., healthcare, pandemic monitoring, etc.). There is also a growing interest in research that examines disparities, peace, justice, and technology ethics. Relatively fewer articles explore sustainability-related issues. Following Ketter et al. (2022), we hope to see more work on sustainability and design in contexts such as smart mobility. Regarding the socio-technical positioning of DSSG research, fewer articles have explored challenges at the institutional level, with greater impetus on people and organizations. This is a microcosm of a broader trend related to relatively fewer articles focusing on technology and institutions (Barley & Orlikowski, 2023).

Our analysis of DSSG in IS and adjacent fields reveals that there are opportunities for researchers to take a more significant thought leadership role regarding social good challenges using data science research genres. Given the broader trends of exponential growth in data science research as a whole (Figure 1) and proportionally less emphasis on social good themes (Figure 2), the opportunity and impetus for IS are perhaps even greater. In the next section, we discuss guidelines for DSSG research in IS.

## 4.2 Guidelines for DSSG Research

### 4.2.1 Beyond Availability Biases – Going the Last Research Mile

In June 2020, the crowd-sourced business review platform Yelp added a feature allowing businesses to label themselves as black-owned. The move spurred a flurry of working papers and publications examining the impact of such a label on sales and/or review ratings for black-owned businesses (Aneja et al., 2023; Babar et al., 2023), as well as work on racial disparities related to borrowing through the paycheck protection program (Evans, 2021; Chernenko & Scharfstein, 2022). In total, between January 2021 and September 2023, we observed an astounding 108 new studies added to the Google Scholar index that examine the impact of the new "black-owned" labeling feature on Yelp (as a natural experiment) or that use the label to explore other types of disparities (e.g., discrimination faced by minority small-business owners). On the one hand, this example shows that the academic research community is committed to exploring social good challenges. It suggests that the lack of earlier research on such questions is not a function of apathy. On the other hand, it also speaks to the academic research communities' over-reliance on readily available, easy-to-collect data sources curated from API/platform-driven environments.

For DSSG to thrive, what is needed are programmatic streams of high-impact research (Ram & Goes, 2021). Characteristics of programmatic research include (Ram & Goes, 2021): (1) that it is thoughtfully and deliberately designed, rather than being opportunistic; (2) that it is geared towards tackling big questions/challenges; (3) and that it often relies on significant (primary) data collection efforts. Let us contrast this with the status quo for data science research. Using the computational design genre as an example, within that genre, data science artifacts are often evaluated and validated based on how well they perform across a set of well-established performance metrics (e.g., accuracy and sensitivity). The importance of such metrics has further amplified in recent years with the rise of data analytics competitions, crowd-sourcing platforms, and leaderboards. While such metrics are important, and in some respects, they constitute the "price of admission" for artifact design, they often fail to consider key downstream implications – humanistic outcomes and societal impact. Some scholars have described this as "going the last research mile…using scientific knowledge and methods to address important unsolved classes of problems for real people with real stakes in the outcomes" (Nunamaker et al., 2015, p. 11).

To illustrate this difference between an article and a programmatic research stream that goes "the last mile," we can look at the important work on computational design in the context of mental health. With the broader availability of social-media data labeled with the authors' mental health status, two review articles on machine learning-based depression detection from social media identified over 110 new articles published on the topic since 2016 (Liu et al., 2022; Malhotra & Jindal, 2022;). However, almost none discuss testing their proposed models/methods in actual field environments. Nor do these studies highlight the efficacy of placing this much emphasis on social media data relative to clinical data settings shown to have less veracity and higher clinical validity for diagnosing mental health conditions (Seyedi et al., 2023). Conversely, the 2020 winner of the INFORMS ISS Design Science Award was a multi-year research stream related to identifying online users battling emotional distress and developing monitoring and decision-support tools for workers in suicide prevention centers (Li et al., 2014; Chau et al., 2020).

It is important to note that in the emerging data science for social good literature, some are setting the bar for social impact even higher. For instance, Cowls et al. (2021) propose five criteria for "AI 4 social good" research that include "projects built and used in the field for at least six months," and that have "documented positive impact." Taylor et al. (2019) highlight the importance of engaging with the communities affected by social good research with their "seven habits of highly successful research in special populations." We recognize that programmatic research may fall along a continuum, and not every stream will meet such stringent criteria. At the very least, we hope that in the early stages of research projects, more conversations will begin with "what's important?" and "what might be impactful?" rather than "what's readily available?"

### 4.2.2 From the Common Task Framework to a Common Good Framework

In recent years, data science has progressed considerably under the common task framework (CTF) (Liberman, 2010; Donoho, 2017). Originally designed for predictive machine learning models at DARPA in the 1980s and later adopted by the IARPA and NIST-sponsored Text REtrieval Conference (TREC) in the 1990s for large-scale evaluation in information retrieval contexts, CTF has been defined as the "quantitative comparison of alternative algorithms on a fixed task" (Liberman, 2010, p. 598). It relies on the construction of publicly available training datasets, mechanisms to evaluate trained models against test data, and a predefined set of key performance indicator metrics (Donoho, 2017). CTF is undoubtedly a driving force behind advancements in state-of-the-art computer vision and natural language processing, resulting in dramatic improvements in an array of downstream tasks related to image recognition and language modeling capabilities. CTF has produced hundreds of thousands of research articles in recent years, fueling the aforementioned explosion of publications and citations pertaining to deep learning.

CTF has drawn attention from various fields. In psychology, it inspired efforts related to reproducibility and replication through the Open Science Framework (osf.io). In the social sciences, it is the basis for the novel experiment design known as behavioral mega-studies (Milkman et al., 2021), which aim to use common data collection to test a large number of hypothesized treatments in one fell swoop (thereby controlling for differences in populations, time, and space). Statisticians attribute the rise of predictive modeling (as a major thrust) within data science to CTF (Donoho, 2017). In IS, the benefits of such common benchmark datasets have been noted (Padmanabhan et al., 2022, p xiii), with researchers invited "to consider what it would take to construct benchmark evaluations in our field."

Despite the many merits and successes of CTF, we cannot help but feel that, in some respects, data science has been led astray. Case in point, in natural language processing, there are well over 100 common tasks for text classification. However, only a few publicly available testbeds and tasks exist for assessing the fairness of text classification models (Abbasi et al., 2021; Guo et al., 2022; Lalor et al., 2022). Access to data remains an impediment to socially good data science research – a point we alluded to in our discussion of availability biases. What we need is a *common good framework* – a collection of social good tasks comprising publicly available data sets and/or access to field research, with established success criteria and guardrails for avoiding unintended consequences of data science (Cowls et al., 2021).

### 4.2.3 Aligning Incentives – Reconsidering Research Productivity and Impact Metrics

A major obstacle to overcoming availability biases and developing common good frameworks is status quo incentive structures for early-career researchers. The current flavor-of-the-month for "publish or perish" emphasizes *where* to publish, and *how many* to publish. On the question of appraising *what* was published, the standard-keepers are undiscerning. This is reflected in our academic productivity rankings (another type of leaderboard with unintended consequences), which rank individuals, departments, colleges, and universities based on the number of publications ($X$) in a pre-defined set of journals ($Y$). Using data science speak, unsurprisingly, this flavor-of-the-month is causing many scholars to treat research as a min-max problem – how to get a desired $X$ in $Y$ while minimizing the maximum effort needed, resulting in minimal viable product research.

Admittedly, the min-max reference might be overly cynical. Nonetheless, any reasonable "accounting of the counting" in research productivity measurement will reveal that if the goal is to focus on impact, the incentive misalignment is very real. We have talked to countless (pun intended) early-career scholars who reference the famous PhD.com comic[5] about the (lack of) "Evolution of Intellectual Freedom" across one's career. Coincidentally, many of the same early-career scholars describe their scientific process as the "need for speed" – getting their research accepted before other researchers studying the same problems, using the same data, with the same methodological perspectives. The silver lining is the renewed emphasis on promoting social good research through special issues at top venues, such as this one and others (Aanestad et al., 2021; Dutta et al., 2023), and impact-oriented research awards[6]. We hope more such publication and recognition opportunities will be available in the coming years.

---

[5] https://phdcomics.com/comics/archive.php?comicid=1436

#### 4.2.4 Large Language Models as a Frontier for Understanding Social Good

These days, no discussion of data science is complete without mentioning large language models (LLMs) such as generative pre-trained transformers (GPTs) (Brown et al., 2020). In our discussion of the DSSG framework, we noted the exciting trend of using LLMs for computational social science research (Ziems, 2023). We discussed how, in their aptly titled paper, "The Geometry of Culture," Kozlowski et al. (2019) trained a word embedding on text from millions of books published over 100 years to perform a historical analysis of the evolution and dynamics of shared understandings of social class. We also described how Garg et al. (2018) used word embeddings to quantify 100 years of gender and ethnic stereotypes. Some computational social science researchers have argued that the far greater levels of specificity associated with algorithms could be useful for probing aspects of human decision-making (generally considered to be less transparent due to a myriad of reporting/disclosure biases) to help detect discrimination in human decision-making (Kleinberg et al., 2020). Terms such as *digital anthropology* – the use of digital technologies within the anthropological methodology (Miller, 2018) – and *cyber archaeology* – the study of cultural cyber artifacts (Zimbra et al., 2010) – have previously been used to connote the confluence of digital and physical worlds as it relates to understanding the human condition, groups, societies, and cultures. Advancements in LLMs have the potential to "revolutionize anthropological research and practice" in business/organizational settings (Artz, 2023), where digital traces such as documents, emails, meeting minutes, transcriptions, etc., capture genres, power structures, organizational routines, norms, culture, and so forth.

Of all the available state-of-the-art tools, methods, and artifacts for data science, we specifically mention LLMs in part due to their ability to capture rich interpersonal patterns across massive corpora, thereby overcoming some of the data/resource accessibility limitations of traditional large-scale longitudinal field data collections. Moving forward, we believe LLMs will have multiple roles to play in the context of DSSG in IS: (1) as a mechanism for overcoming availability biases in research; (2) as a computational social science tool for studying social good challenges by examining patterns codified in LLMs pretrained and/or fine-tuned on relevant individual, organizational, or institutional corpora; (3) as a source of codified stereotypes and other related disparities in modern generative AI; (4) as the basis for a new wave of computationally intensive theory building and computational design artifacts related to social good challenges.

## 5 Introduction to the Special Issue

IS scholars are interested in using their research to address various societal challenges and make a difference. This can be seen by the many inquiries about and the responses to the Call for Papers on this special issue announced in August 2020. We want to thank the 44 groups of authors who submitted their extended abstracts in November 2020 for comments and feedback. In April 2021, based on our feedback related to abstracts, we received 20 full-paper submissions. A team of 34 researchers with extensive knowledge, expertise, and research records in data science reviewed these submissions. Their valuable and insightful comments provided authors with ideas and constructive instructions to improve their manuscripts. The dedication of these reviewers helped us immensely during the review process. Without their outstanding contributions, for which we are grateful, we would not have been able to publish this special issue. After four rounds of extensive review and author revisions, we accepted four articles for this special issue. These four articles represent a diverse range of DSSG genres, challenges, and socio-technical environments. See Figure 6.

The article entitled "ShowCase: A Data-Driven Dashboard for Federal Criminal Sentencing," presents a study in the technology and institution's socio-technical environment. It seeks to address the possible inequality and disproportionate sentencing of minorities in the legal and justice context. Although this article is mostly in the computational design genre, it is also highly related to computational social science. The design artifact, a data-driven dashboard named ShowCase, is grounded in theories from the organizations and penal justice literature. The dashboard can help judges make fairer and more objective decisions by integrating a variety of data points. This research has the potential to promote fairness, objectivity, and transparency in the criminal justice system.

---

[6] For instance, the ISS Bapna-Ghose Social Justice Award: https://www.informs.org/Recognizing-Excellence/Community-Prizes/Information-Systems-Society/ISS-Bapna-Ghose-Social-Justice-Best-Paper-Award

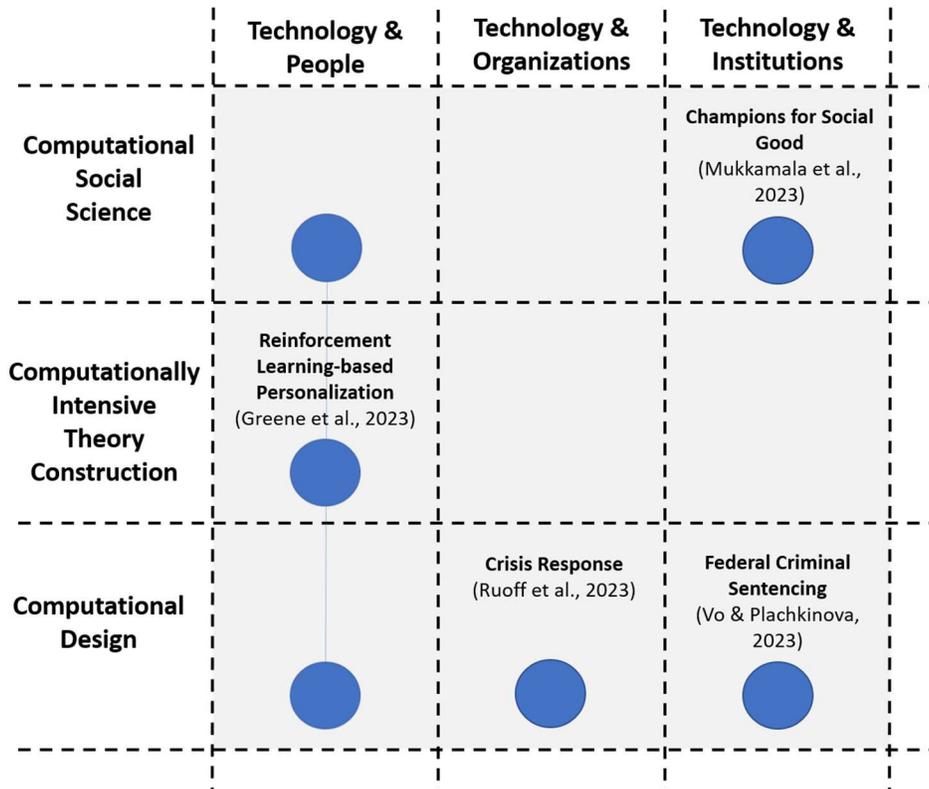

**Figure 6:** Grouping of Four Special Issue Articles Based on our DSSG-IS Framework

The research presented in the article, entitled "Designing Conversational Dashboards for Effective Use in Crisis Response," is intended to enhance community well-being by helping government agencies and health organizations improve their responses to crises such as the COVID-19 pandemic. Similar to the first article, this computational design research employs the design science methodology. It develops a conversational dashboard that allows users to use natural languages to interact with the system. The evaluation of the dashboard shows that users can find their needed information more effectively and efficiently – important outcomes when sense-making during crisis response.

The third article, entitled "Champions for Social Good: How can we Discover Social Sentiment and Attitude-driven Patterns in Prosocial Communication?" demonstrates how DSSG research can contribute to the efforts of enhancing harmony and peace at the society level. Motivated by the social media strategy of the United Nations High Commissioner on Refugees (UNHCR), this research studies how the Twitter communication of high-profile prosocial influencers or champions for change might have influenced their followers and audiences. Using a Twitter sentiment and attitude corpus, an analytics framework is proposed, which consists of machine learning and natural language processing and is used to test the impact of different types of refugee-related emergencies and champion influencers, on patterns observed in social communication. This research showcases the contributions of data science research to prosocial policies regarding refugee crisis awareness in a broader institutional context.

The article "Taking the Person Seriously: Ethically-aware IS Research in the Era of Reinforcement Learning-based Personalization" is positioned more as a research commentary. It posits that the development of reinforcement learning techniques, which have been increasingly employed in personalization and adaptive control of individuals' environments, may endanger persons and societies at scale. Five emergent features of this new personalization paradigm are identified, and their potential dangers are discussed, including diminished personal autonomy, social and political instability, mass surveillance and social control, and privacy invasions, among other concerns. Because these potential issues cannot be addressed adequately by current data protection laws and regulations, this article proposes three directions for ethically-aware reinforcement learning-based personalization research uniquely suited to the strengths of IS researchers across the socio-technical spectrum. Although this article does not directly fit into the three data science genres explicated in our DSSG-IS framework, per se, we believe it has far-reaching implications directly related to all three genres. For instance, the article talks about ethical design, offers lexical framing for studying the social challenges of reinforcement learning, and provides a bevy of problem contexts where computational social science may illuminate the size and scope of issues created by reinforcement learning.

These four articles add knowledge and insights to the DSSG literature and identify potential research directions for addressing various societal challenges and issues. They employ different research genres and focus on unique social good challenges in distinct socio-technical environments. We hope readers find them informative and helpful for their future DSSG research. We also hope that IS data science research can reverse the alarming trend we observed across all data science research over the past 30-plus years in which social good challenges are garnering proportionately less attention with each passing day.